\documentstyle{elsartwb}

\input psfig
\input epsf

\begin{document}

\begin{frontmatter}

\title{Vanishing dynamical quark mass at zero virtuality? \thanksref{grants}}
\thanks[grants]{Research supported in part by
        the Polish State Committee for
Scientific Research grant 2P03B-080-12,
the Russian Foundation of Basic Research grant 01-02-16431,
INTAS-00-00366,
and by the
Bogoliubov-Infeld programm.}
\thanks[emails]{%
\hspace{0mm} E-mail: dorokhov@thsun1.jinr.ru,
broniows@solaris.ifj.edu.pl}
\author[Dubna]{Alexander E. Dorokhov} and
\author[INP]{Wojciech Broniowski}\address[Dubna]
{Joint Institute for Nuclear Research, Bogoliubov Laboratory of
Theoretical Physics,
114980, Moscow region, Dubna, Russia }
\address[INP]{H. Niewodnicza\'nski Institute of Nuclear
Physics,
         PL-31342 Krak\'ow, Poland}

\begin{abstract}
We show that the dynamical quark mass in effective nonlocal models can vanish
at zero virtuality of the quark as $M(p^2)\propto p^2$.
Our arguments follow from the constrained-instanton
model of the QCD vacuum and from QCD sum rules calculations
with nonlocal condensates.
The discussed models also lead to analyticity of $M(p^2)$ in the vicinity of zero.
\end{abstract}


\end{frontmatter}

\vspace{-7mm}


The concept of dynamical quark mass generated by the spontaneous breaking of
chiral symmetry has proven very useful in the description of low-energy
phenomena. This mass enters the quark propagator, $
S_{F}^{-1}(p)=Z(p^{2})[\FMslash{p}+M(p^{2})]$, and, in general, depends on the
(Euclidean) momentum $p$. Specific forms of $M(p^{2})$ are obtained in
instanton models \cite{Shuryak96} or via the Schwinger-Dyson resummation
techniques \cite{RobWil}. The quark mass in its own is not observable and
does not bear physical significance. Moreover, it depends on the choice of
the gluon field gauge, hence is not unique. Also, the region where,
{\em e.g.,} the instanton models dominate is at some intermediate values of $p^{2}$, not
too high, where the model predicts a too fast dropping of the mass in
the perturbative region, and not too low,
since certainly the infrared physics (confinement)
is not being incorporated properly.
Nevertheless, the mentioned popular approaches have founded a typical image
of $M(p^{2})$ in the Euclidean space: it is {\em non-zero} at $p^{2}=0$,
drops monotonically with increasing $p^{2}$, and asymptotically reaches the
current quark mass value, $m_{c}.$\footnote{
Throughout this paper we are working in the strict chiral limit of $m_{c}=0$.
}

In this paper we argue that this conventional picture
need not be the case, and that we may well have
\begin{equation}
M(p^{2}=0)=0.  \label{zero}
\end{equation}
As $p^{2}$ is increased, $M(p^{2})$ grows, arrives at maximum, and then
decreases to reach $m_{c}$ asymptotically. In fact, we will present
arguments that the situation (\ref{zero}) can be more natural than the
conventional case of $M(p^{2}=0)>0$. One class of arguments comes from
QCD sum rules with nonlocal condensates, another one from the so-called ``constrained''
instantons, where the large-distance asymptotics is improved by an
additional cut-off function in the instanton size.

We will also show that although Eq. (\ref{zero}) looks, at first glance,
unusual, it does not lead to pathologies: the chiral symmetry is broken, and
all results enforced by this fact hold: the quark condensate has a non-zero
value, the Gell-Mann-Oaks-Renner relation is satisfied, as are other
relations coming from symmetries, anomalies are preserved, {\em etc.}

We begin by presenting a calculation in framework of the constrained
instanton model, where Eq. (\ref{zero}) holds. We use for simplicity the
``constrained'' quark zero mode in the singular gauge and introduce
the following ansatz for the profile function \cite{Espin90}:
\begin{equation}
\psi _{sing}^{\pm }(x)=\sqrt{2}\varphi _{sing}(x)\frac{\widehat{x}}{\left|
x\right| }\chi ^{\pm },\qquad \varphi _{sing}(x)=\frac{\overline{\rho }
(x^{2})}{\pi (x^{2}+\overline{\rho }^{2}(x^{2}))^{3/2}},  \label{PsiReg}
\end{equation}
where $\chi $ is a color Dirac spinor given by $\chi ^{\pm }\overline{\chi }
^{\pm }=\left( \gamma _{\mu }\gamma _{\nu }/16\right) \left( 1\pm \gamma
_{5}\right) /2\tau _{\mu }^{\pm }\tau _{\nu }^{\mp }$ and $\tau _{\nu }^{\pm
}=\left( \mp i,\overrightarrow{\tau }\right) $, with the upper (lower) signs
corresponding to solutions in the instanton (anti-instanton) field. The
standard instanton solution is obtained with the constant for $\overline{
\rho }^{2}(x^{2})=\rho ^{2}$. For the constrained instantons one uses
exponentially-decreasing functions $\overline{\rho }^{2}(x),$ normalized as
$\overline{\rho }^{2}(0)=\rho ^{2}$. We shell use the form \cite{DEMM99}
\begin{equation}
\overline{\rho }^{2}(x^{2})=\frac{2}{\Gamma (1/3)3^{1/3}}\left( \frac{
\rho }{R}\right) ^{2}x^{2}K_{4/3}\left( \frac{2}{3}\left( \frac{x^{2}}{R^{2}}
\right) ^{3/4}\right) ,  \label{rhoX}
\end{equation}
where $K_{\nu }\left( z\right) $ is the modified Bessel function. The
specific feature of the constrained instanton is that at small distances it
is close to the standard instanton profile of size $\rho$, and at large distances it has
exponentially-decreasing asymptotics governed by a large-scale parameter $R$,
such as $\rho<R$. These shapes
are motivated by considering modifications of the
instanton in the background field of large-scale vacuum fluctuations
\cite{DEMM99}. The constrained instanton profile, as opposed to the
unconstrained one, provides the correct large-distance
asymptotics of the quark and gluon field correlators \cite{DEMM99}.

Let us define the 4-dimensional Fourier transform of the quark zero mode profile
$\widetilde{\varphi }_{sing}(p)$. The explicit form of the Fourier transform
of the standard (unconstrained) zero mode in the singular gauge is
well known,
\begin{equation}
\widetilde{\varphi }_{sing}^{{\rm I}}(p^{2})=\pi \rho ^{2}\frac{d}{dz}\left[
I_{1}(z)K_{1}(z)-I_{0}(z)K_{0}(z)\right] _{z=\rho p/2},  \label{Phi}
\end{equation}
and has the asymptotics
\begin{equation}
\widetilde{\varphi }^{{\rm I}}(p^{2})=\left\{
\begin{array}{c}
\frac{2\pi \rho }{p}\qquad {\rm as}\qquad p^{2}\rightarrow 0, \\
\frac{12\pi }{\rho ^{2}p^{4}}\qquad {\rm as}\qquad p^{2}\rightarrow \infty.
\end{array}
\right.  \label{Phi(p)_Reg}
\end{equation}
The constrained zero mode has the same large-$p^{2}$ asymptotics,
but it goes to a constant at $p^{2}\rightarrow 0$
\begin{equation}
\widetilde{\varphi }^{{\rm CI}}(p^{2})=\left\{
\begin{array}{c}
\pi ^{2}I_{{\rm CI}}\qquad {\rm as}\qquad p^{2}\rightarrow 0, \\
\frac{12\pi }{\rho ^{2}p^{4}}\qquad {\rm as}\qquad p^{2}\rightarrow \infty ,
\end{array}
\right.  \label{Phi(p)_RegCI}
\end{equation}
where $I_{{\rm CI}}=\int_{0}^{\infty }du\; u\varphi _{{\rm CI}}(u)$.
The quark mass generated by the  one-instanton background is equal to \cite{CDG78}
\begin{equation}
M(p^{2})=Cp^{2}\widetilde{\varphi }^{2}(p^{2}),  \label{Mp2}
\end{equation}
where the constant $C>0$ is determined from the gap equation \cite{CDG78,DP86}
\begin{equation}
\int \frac{d^{4}k}{(2\pi )^{4}}\frac{M^{2}(k^{2})}{k^{2}+M^{2}(k^{2})}=\frac{
n}{4N_{c}},  \label{Gap2}
\end{equation}
with $n = 0.0016$ GeV$^4$ \cite{Shuryak96} denoting the
instanton density. From the asymptotic behavior (\ref{Phi(p)_Reg})
and (\ref{Phi(p)_RegCI}) we see immediately, that for the
standard zero mode $M_{I}(0) > 0$, whereas constrained zero modes give $
M_{CI}(0)=0$. In Fig. 1 we show the quark mass function for the constrained
and unconstrained zero modes. As advocated above, the constrained
zero modes lead to (\ref{zero}). The momentum at which $M(p^2)$
starts dropping as $p^2$ is being decreased is controlled by the large-scale parameter $R$.
When $R$ is increased, the constrained profile tends to the standard zero mode solution
everywhere except the point $p^2=0$.

\begin{figure}[tb]
\vspace{0mm} \epsfxsize = 9.5 cm \centerline{\epsfbox{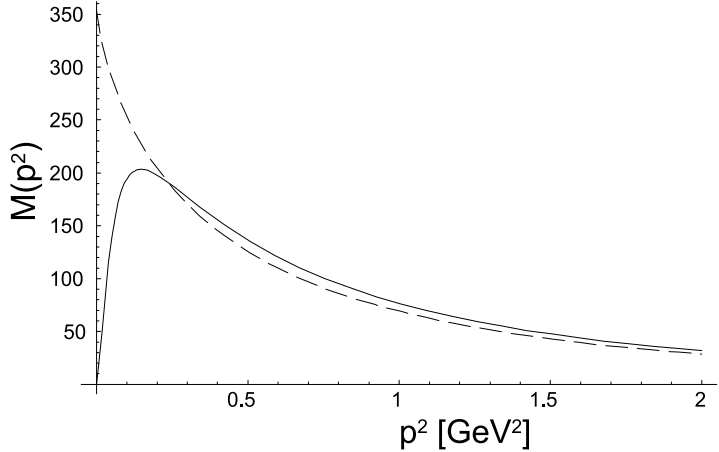}} \vspace{0mm}
\label{fig1}
\caption{Dynamical quark mass as a function of the square of the
Euclidean momentum. The dashed line corresponds to the standard zero mode
solution of Eq. (\protect\ref{Phi}), while the solid line corresponds
to the constrained
zero mode profile, Eq. (\protect\ref{rhoX}). The parameters
are $\rho=1.7{\rm GeV}^{-1}$ and $R=3\rho$.}
\end{figure}

Let us pass to another approach hinting to Eq. (\ref{zero}), namely QCD sum
rules with nonlocal condensates
$\tilde Q(x^2)\equiv\left\langle \overline{q}(x)P\exp{[ig\int_0^x\ dzA(z)]}
q(0)\right\rangle $, {\em etc} \cite{MihRad92}.
The local quark condensates, $<\bar{q}q>$,
$<\bar{q}(D^{2})^{n}q>$, $n=1,2,...$ appear as expansion coefficients of the
correlator $\tilde Q(x^2)$ in the variable $x^2/4$.
Now, we make a crucial assumption needed for our argumentation:
the quark condensate and its moments can be
identified with  the expressions obtained at the one-quark-loop level in the
effective quark model. This is valid in the large-$N_{c}$ limit.
We have therefore (for the single-flavor)\footnote{
For simplicity, we are neglecting the effects of the quark wave-function renormalization by
setting $Z(q^{2})=1$ in Eq. (\ref{Gap2}). This amounts to neglecting
possible vector interactions, as is the case of the instanton
calculations.}
\begin{equation}
\langle \bar{q}q\rangle =-4N_{c}\int \frac{d^{4}p}{(2\pi )^{4}}\frac{M(p^{2})
}{p^{2}+M^{2}(p^{2})},  \label{QQcond}
\end{equation}
and
\begin{equation}
\lambda_q^{2n}\equiv
\frac{\langle \bar{q}D^{2n}q\rangle }{\langle \bar{q}q\rangle }=-\frac{4N_{c}
}{\langle \bar{q}q\rangle }\int \frac{d^{4}p}{(2\pi )^{4}}p^{2n}\frac{
M(p^{2})}{p^{2}+M^{2}(p^{2})}.  \label{D2n_M}
\end{equation}
can be found\footnote{
Here and below we silently assume that nonlocal condensates are
defined in gauge-invariant way by putting in
appropriate definitions the path-ordered Schwinger
phase factor. In QCD sum rules this factor becomes unity throughout the
use of the Fock-Schwinger gauge. A similar situation occurs in the
instanton model, where we have to transform the fields from standard
(singular or regular) gauges to the axial one \cite{DEM97}. This results in modification
of Eq. (\ref{Mp2}). We omit these complications in
the present work, since the influence of the phase factor is not important for the
qualitative analysis of the dynamical quark mass.}
\begin{equation}
Q(p^{2})=2\frac{M(p^{2})}{p^{2}+M^{2}(p^{2})},  \label{QofM}
\end{equation}
where $Q(p^{2})$ is proportional to the Fourier transformation of the scalar quark condensate $\tilde Q(x^2)$.
In QCD sum rules one typically makes a hypothesis for $Q(p^{2})$.
For instance, one of the ansatze successfully applied in QCD sum rules \cite{MihRad92} is
\begin{equation}
Q(p^{2})=-\frac{\langle \bar{q}q\rangle }{2N_{c}}\frac{16\pi ^{2}
}{\Lambda ^{4}}\exp (-\frac{p^{2}}{\Lambda ^{2}}),  \label{qbmr}
\end{equation}
with average quark virtuality in QCD vacuum being
\begin{equation}
\lambda_q^2=2\Lambda^2.
\label{L2q}\end{equation}
Then, the moments of the quark condensate become
parameterized in a simple way. Thus, in QCD sum rules its the nonlocal quark
condensate $Q(p^{2})$ which is the basic quantity, which is in a sense
opposite to quark models or the instanton approach, where we start with
$M(p^{2})$. The ansatz (\ref{qbmr}) is compatible with recent lattice measurements of scalar quark
correlator, $\tilde Q(x^{2})$, on the lattice \cite{DiGi99,Meggi00}. The dynamical quark mass $M(p^{2})$ is related to the
nonlocal quark condensate $Q(p^{2})$ via Eq. (\ref{QofM}).

We shall now explore the consequences of Eq. (\ref{QofM}) and try to
invert it in order to obtain the quark mass
as a function of the condensate (assumed to
be known from the QCD sum rules \cite{MihRad92} or lattice simulations \cite{DiGi99}).
A few implications of (\ref{QofM}) are immediate. We can see that due
to the rapid decrease of the profiles $Q(p^{2})$ and $M(p^{2})$ at large $p^{2}$
the relation becomes linear asymptotically:
\begin{equation}
M(p^{2})=\frac{1}{2}p^{2}Q(p^{2})\qquad {\rm at\ large}\quad p^{2}.
\label{MofQ_largeK}
\end{equation}
This is the dilute instanton medium regime, where the single-instanton
effects dominate. Next, from (\ref{QofM}) it follows that for real valued $M(p^2)$
\begin{equation}
\sqrt{p^{2}}Q(p^{2})\leq 1  \label{kQ_bound}
\end{equation}
for any value of $p^{2}>0$. Thus, in general, one can consider three different
situations. First, if the model profile function for the nonlocal condensate
$Q(p^{2})$ violates the bound (\ref{kQ_bound}), then one cannot find a
quark-model representation for the results in terms of real valued
functions.
The other two possibilities, discussed by us below,
correspond to the case when the bound
(\ref{kQ_bound}) is saturated at some points $\sqrt{p^{2}}\equiv m_{i}$, or,
finally, to the case of the strict inequality for all momenta, $\sqrt{p^{2}}
Q(p^{2})<1$. Let us first consider the last possibility, where we find
\begin{equation}
M(p^{2})=\frac{1-\sqrt{1-p^{2}Q^{2}(p^{2})}}{Q(p^{2})}.  \label{MofQ1}
\end{equation}
The sign of the root is chosen in such way that the large-$p^{2}$
asymptotics (\ref{MofQ_largeK}) is satisfied.
The feature of this solution is that $M(p^{2}=0)=0$.
Indeed, at small $p^{2}$ we get
\begin{equation}
M(p^{2})=\frac{1}{2}Q(0)p^{2}+O(p^{4}).
\end{equation}

In the case where the bound Eq. (\ref{kQ_bound}) is saturated at real valued points
$\sqrt{p^{2}}\equiv m_{i}$, the solution can {\em smoothly}
jump from one branch to another. We illustrate this for
the case of a single point, $m_{0}$. At this point we have the identity
\begin{equation}
M(m_{0}^{2})=m_{0}.  \label{M(m)=m}
\end{equation}
In this case the inverse of (\ref{QofM}) is
\begin{equation}
M(p^{2})=\frac{1-{\rm sgn}\left( p^{2}-m_{0}^{2}\right)
\sqrt{1-p^{2}Q^{2}(p^{2})}}{Q(p^{2})}.  \label{MofQ}
\end{equation}
The sign function flips the solution from one branch to another at the point
$p^{2}=m_{0}^{2}$. This prescription ensures the continuity of $M(p^{2})$
with all its derivatives. Note that now, due to branch-switching, we have
$M(0)=\frac{2}{Q(0)}>0.$

We illustrate the above considerations in the model (\ref{qbmr}). The
condition (\ref{M(m)=m}) occurs for
\begin{equation}
\Lambda_{\rm crit}^{3}=-\frac{8\pi ^{2}}{\sqrt{2e}}\frac{\langle \bar{q}q\rangle }{%
N_{c}},  \label{tune}
\end{equation}
which numerically gives $\Lambda_{\rm crit}=505{\rm MeV}$
if one fixes $\langle \bar{q}q\rangle = -(225{\rm MeV})^3$.
At $\Lambda > \Lambda_{\rm crit}$ we have only one
branch. At $\Lambda = \Lambda_{\rm crit}$ we
have the branch-switching, according
to (\ref{MofQ}). The situation is depicted in Fig. 2.

\begin{figure}[tb]
\vspace{0mm} \epsfxsize = 9.5 cm \centerline{\epsfbox{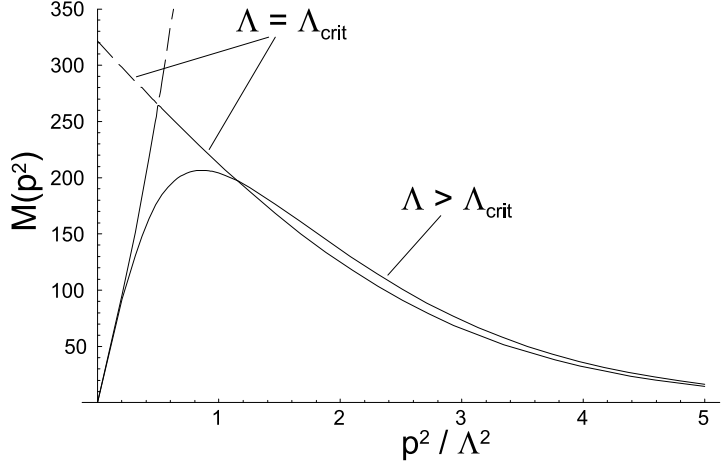}} \vspace{0mm}
\label{fig2}
\caption{Dynamical quark mass for the nonlocal condensate given
by Eq. (\protect\ref{qbmr}), plotted as
a function of the square of the Euclidean
momentum expressed in units of the cut-off parameter $\Lambda$. The curve
labeled $\Lambda > \Lambda_{\rm crit}$
satisfying condition (\protect\ref{zero}) corresponds to
$\Lambda = 554{\rm MeV}$, which fits the value of
the pion decay constant in the chiral limit, $F_\pi=86$ MeV. The
curve labeled $\Lambda = \Lambda_{\rm crit}$ displays the branch-switching. At
high momenta the solution is given by the solid line, and at low momenta by the
dashed line.}
\end{figure}

One determines the values of model parameters, such as $\Lambda$, by
fitting observables. In quark models one typically fits the pion decay
constant, $F_{\pi }$ to its experimental value, which in the chiral limit
equals $86$ MeV  \cite{GL84}. The expression
for $F_{\pi }$ is \cite{DP86,Birse95}
\begin{equation}
F_{\pi }^{2}=\frac{N_{c}}{4\pi ^{2}}\int\limits_{0}^{\infty }du\ u\frac{%
M(u)^{2}-uM(u)M^{\prime }(u)+u^{2}M^{\prime }(u)^{2}}{\left(
u+M(u)^{2}\right) ^{2}},  \label{Fpi2_M}
\end{equation}
or it may be rewritten in terms of $Q$ as:
\begin{equation}
F_{\pi }^{2}=\frac{N_{c}}{64\pi ^{2}}\int\limits_{0}^{\infty }du\ u\frac{
4Q(u)^{2}+4u Q(u)Q^{\prime }(u)+4u^{2}Q^{\prime }(u)^{2}-3uQ(u)^{4}}
{1-4uQ(u)^{2}}.  \label{Fpi2_Q}
\end{equation}
Above, $M^{\prime }(u)=\frac{d}{du}M(u)$ and $Q^{\prime }(u)=\frac{d}{du}Q(u)$.
Figure 3 shows $F_\pi$ evaluated with (\ref{qbmr}), plotted as a
function of $\Lambda$. We note
that the experimental value of $F_\pi$ favors $\Lambda\approx 0.55$ GeV, comfortably
above the critical value, such that
indeed we have a one-branch situation and, consequently, (\ref{zero}).
The average quark virtuality (\ref{L2q}) is estimated as  $\lambda_q^2=0.6$ GeV$^2$
and fits the value obtained in the QCD sum rules $\lambda_q^2=0.5\pm0.1$ GeV$^2$ \cite{BM95}.
Another interesting feature of ansatz (\ref{qbmr}) is that the quark propagator does exhibit
analytical confinement of quarks. This is because the equation $p^2=M^2(-p^2)$ has no real solution at
positive (Minkowskian) $p^2$.

\begin{figure}[tb]
\vspace{0mm} \epsfxsize = 9.5 cm \centerline{\epsfbox{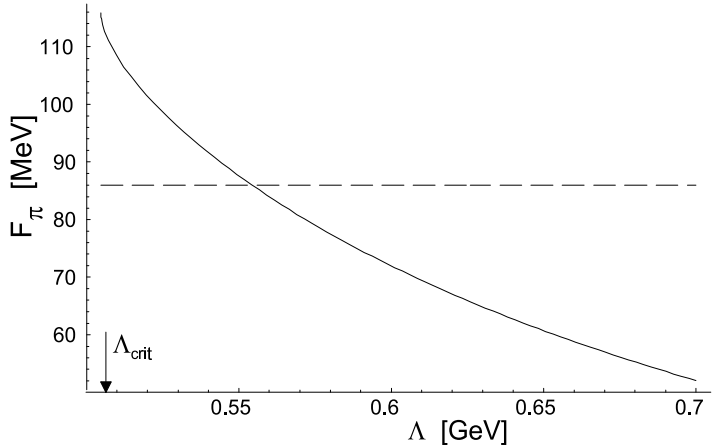}} \vspace{0mm}
\label{fig3}
\caption{$F_{\pi}$ as a function of $\Lambda$ (solid line) and the experimental
value of 86MeV (dashed line). }
\end{figure}

The important feature of effective models is the compliance to symmetries,
in particular to the chiral symmetry. The consistency of quark models with
nonlocal regulators with symmetries has been shown in Ref. \cite{Birse95}
and we are not going to repeat it here. Since the
proofs never use the explicit form of $M(p^2)$, the results are true for any
mass function, and thus also hold for functions with the property (\ref{zero}).

Finally, we examine the analytic properties of the pion propagator. The
purpose here is to check whether the vanishing quark mass does not induce an
infrared cut, which would be unphysical. In nonlocal quark models the inverse pion
propagators is given by \cite{Birse95,Birse98,ripka:book,basz}
\begin{eqnarray}
&& K_{\pi }^{-1}(q) = {\rm const}\int \frac{d^{4}p}{(2\pi )^{4}}\times
\label{Kpi} \\
&& \left[ \frac{M(p+\frac{q}{2})M(p-\frac{q}{2})\left( p^{2}-\frac{q^{2}}{4}%
+M(p+\frac{q}{2})M(p-\frac{q}{2})\right) }{\left( {(p+\frac{q}{2})}^{2}+M(p+{%
\frac{q}{2}})^{2}\right) \left( {(p-\frac{q}{2})}^{2}+M(p-\frac{q}{2}%
)^{2}\right) }-\frac{M(p)^{2}}{{p}^{2}+M(p)^{2}}\right] \;.  \nonumber
\end{eqnarray}
It is easy to verify that the mass factors $M(p+\frac{q}{2})M(p-\frac{q}{2})$
in the numerator cancel the infrared divergence from the denominator, and as
a result the pion propagators has no cuts at low $q$. Its analytic structure
in the low-energy region consists of the pole at $q^{2}=0$, as it should be.
If it were not for the presence of $M(p+\frac{q}{2})M(p-\frac{q}{2})$ in the
numerator, expression (\ref{Kpi}) would have developed a cut in the $q^{2}$
variable reaching from $0$ to $-\infty $. Analogous result holds for the
$\sigma $-field propagator.

We conclude that the scenario of a vanishing dynamical quark mass at zero
quark virtuality is possible, as well as free of pathologies. In forthcoming
work we are going to study the processes that are sensitive to infrared
behaviour of nolocal nonperturbative propagator. In particular, in \cite{DoLaurPLB00}
it was shown that the pion light-cone wave function in the middle region is
related to derivative of the vacuum nonlocality in the infrared region. Finally, the models
discussed in this paper lead to, in addition to
Eq. (\ref{zero}), {\em analyticity} of $M(p^2)$ in the vicinity of zero.
This is not the case of, {\em e.g.}, standard zero modes, where
the function develops a cut at $p^2=0$, extending along the whole
Minkowski region. This nonanalyticity becomes an obstacle in soliton
calculations in such models \cite{nls}, where other regulators must be used,
such that they can be continued to the Minkowski momenta near the origin.
Models discussed in this paper are free of that problem.

The authors are thankful to A.P. Bakulev, G.V. Efimov, N.I. Kochelev, S.V. Mikhailov,
S.N. Nedelko, M.K. Volkov for discussions.

\end{document}